# Energy Aware , Scalable , K- Hop Based Cluster Formation In MANET.

By
**Priyanka Chatterjee (07/CSE/18)**
**Nikhil Agarwal (07/CSE/62)**

Under the guidance of Faculty
**Asst. Prof. Tanmay De**
**Asst. Prof. Prasenjit Chowdhury**

**COMPUTER SCIENCE & ENGINEERING**

In Partial Fulfilment of the Requirements
For the Degree of

**BACHELOR OF TECHNOLOGY**

## National Institute of Technology, Durgapur

2011

## ACKNOWLEDGEMENTS

We wish to express my sincere gratitude to our thesis advisors Asst. Prof. Tanmay De and Asst. Prof. Prasenjit Chowdhury of Computer Science & Engineering Department of NIT Durgapur , for guiding us through every step of the thesis and providing us direction and insight on numerous occasions during the course of this work. I would also like to thank our friend , Miss. Nevadita Chatterjee, working in Aricent  Technologies for her constant support and appreciation. We are indebted to our parents for all the encouragement. May I take this opportunity to thank our seniors Sandeep, Mtech student for supporting me in my effort to complete this thesis.

# TABLE OF CONTENTS

# ABSTRACT

The study of Mobile Ad-hoc Network remains attractive due to the desire to achieve better performance and scalability. MANETs are distributed systems consisting of mobile hosts that are connected by multi-hop wireless links. Such systems are self organized and facilitate communication in the network without any centralized administration. MANETs exhibit battery power constraint and suffer scalability issues therefore cluster formation is expensive. This is due to the large number of messages passed during the process of cluster formation.

Clustering has evolved as an imperative research domain that enhances system performance such as throughput and delay in Mobile Ad hoc Networks (MANETs) in the presence of both mobility and a large number of mobile terminals.In this thesis, we present a clustering scheme that minimizes message overhead and congestion for cluster formation and maintenance. The algorithm is devised to be independent of the MANET Routing algorithm. Depending upon the context, the clustering algorithm may be implemented in the routing or in higher layers. The dynamic formation of clusters helps reduce data packet overhead, node complexity and power consumption, The simulation has been performed in ns-2. The simulation shows that the number of clusters formed is in proportion with the number of nodes in MANET.

# 1. INTRODUCTION

## 1.1 MANET:

Wireless technologies have become popular as they exhibit ubiquitous features, fulfilling the demand of network communication anywhere, at anytime. Since portable devices like laptop computers, personal digital assistance (PDAs) and mobile phones require fixed infrastructure such as access point or base stations, therefore they need an access to a static network to support their mobile device services. To provide a solution to this problem Mobile Ad hoc Networks (MANETs) have evolved.

. An ad hoc network is a multihop wireless communication network supporting mobile users without any existing infrastructure. To become commercially successful, the technology must allow networks to support many users. A complication is that addressing and routing in ad hoc networks does not scale up as easily as in the Internet. By introducing hierarchical addresses to ad hoc networks, we can effectively address this complication.

MANETs are autonomous systems consisting of mobile hosts that are connected by multi-hop wireless links. MANETs are decentralized networks that develop through self organization. MANETs are formed by a group of nodes that can transmit, receive and relay data among themselves. In mobile ad hoc network there is no fixed infrastructure therefore the mobile hosts communicate over multihop wireless links. These are often called infrastructure-less networking since the mobile nodes in the network dynamically establish routing paths between themselves.

Dynamic routing protocol has gained substantial importance in the design of ad hoc networks such that it can find routes between two communicating mobile nodes efficiently. This is to resist high degree of mobility that changes the network topology. Flat routing schemes have scalability issues associated with them as they impregnate the network by large amount of information flow within the network. In mobile scenario, nodes possess variable computational and communication power with varying resources. This engenders a hierarchical organization of the network topology and motivates the need for cluster based solution to ensure more efficient resource utilization.

In MANETs the nodes communicate over reliable wireless links within the transmission range of each other. In large MANETs, if two hosts are not within the communication range of each other they communicate if other hosts lying in between are willing to forward packets for them. Therefore every node participates in multi-hop routing to reach all nodes in the network. Flooding is a cause of routing in MANETs which leads to degradation of the efficient use of bandwidth and depletes battery-power of nodes. Hence, many clustering schemes have been proposed where the nodes in a network are divided into clusters. Hierarchical routing is adopted, where a standard proactive or reactive protocol can be used within the cluster but other protocols are used for inter-cluster routing. In the dominating set based clustering schemes routing is done based on a set of dominating nodes (DS). These schemes are quite expensive and not very scalable when the number of nodes in the MANET are large and are moving at high speeds.

Topologies in MANET are random, multihop and dynamic. Moreover, there is scarcity of link bandwidth and transmission power of mobile nodes. Therefore, MANET faces major challenges concerning routing, scalability and management functions.

## 1.2 CHALLENGES IN MANET

- Autonomous- No centralized administration entity is available to manage the operation of the different mobile nodes.

- Dynamic topology- Nodes are mobile and can be connected dynamically in an arbitrary manner. Links of the network vary timely and are based on the proximity of one node to another node.

- Device discovery- Identifying relevant newly moved in nodes and informing about their existence need dynamic update to facilitate automatic optimal route selection.

- Bandwidth optimization- Wireless links have significantly lower capacity than the wired links.

- Limited resources -Mobile nodes rely on battery power, which is a scarce resource. Also storage capacity and power are severely limited.

- Scalability- Scalability can be broadly defined as whether the network is able to provide an acceptable level of service even in the presence of a large number of nodes.

- Mobility: Mobility leads to route changes and packet losses in MANETs.

- Limited physical security- Mobility implies higher security risks such as peer-to-peer network architecture or a shared wireless medium accessible to both legitimate network users and malicious attackers. Eavesdropping, spoofing and denial-of-service attacks should be considered.

- Infrastructure-less and self operated- Self healing feature demands MANET should realign itself to blanket any node moving out of its range.

- Poor Transmission Quality- This is an inherent problem of wireless communication caused by several error sources that result in degradation of the received signal.

- Ad hoc addressing- Challenges in standard addressing scheme to be implemented.

## 1.3 APPLICATIONS OF MANET

- Military - Rapidly deployable battle-site networks, sensor fields, unmanned aerial vehicles.
- Disaster Management - Disaster relief teams that can not rely on existing infrastructure.

- Neighbourhood area Networks – Sharable internet access in high density urban settings.

- Impromptu communications among groups of people

- Automobile communications.

## 1.4 TOPOLOGY CONTROL

This deals with the problem of maintaining a connected topology among the nodes in ad hoc networks. This covers power control and hierarchical topology organization. In power control network connectivity is ensured by altering the power of each node in order to balance one-hop neighbour connectivity whereas hierarchical topology control is an approach referred to as clustering.

## 1.5 ROUTING PROTOCOLS IN MANET

- Proactive Protocols
    - Destination Sequenced Distance Vector (DSDV).
- Reactive Protocols
    - Dynamic Source Routing (DSR).
    - Ad hoc  On demand Distance Vector Routing (AODV)
    - Temporally Ordered Routing Algorithms. (TORA).
- Hybrid
    - Zone Routing
- Hierarchical
    - Cluster based
    - Connected -dominating -set -based.

## 1.6 CLUSTERING

Clustering provides a method to build and maintain hierarchical addresses in ad hoc networks. Clustering refers to a technique in which MANET is divided into different virtual group; generally nodes which are geographically adjacent are allocated into the same cluster driven by set of protocols based on behaviours and characteristics of the node. This enables the network to become manageable. Various clustering techniques allow fast connection and also better routing and topology management of MANET.

## 1.6.1 MOTIVATION FOR CLUSTERING IN MANET

It has been shown that cluster architecture guarantees basic performance achievement in a MANET with a large number of mobile terminals. A cluster structure, as an effective topology control means, provides at least three benefits. First, a cluster structure facilitates the spatial reuse of resources to increase the system capacity .With the non-overlapping multicluster structure, two clusters may deploy the same frequency or code set if they are not neighbouring clusters. Also, a cluster can better coordinate its transmission events with the help of a special mobile node, such as a cluster head, residing in it. This can save much resources used for retransmission resulting from reduced transmission collision. The second benefit is in routing, because the set of clusterheads and cluster gateways can normally form a virtual backbone for inter-cluster routing, and thus the generation and spreading of routing information can be restricted in this set of nodes. Last, a cluster structure makes an ad hoc network appear smaller and more stable in the view of each mobile terminal. When a mobile node changes its attaching cluster, only mobile nodes residing in the corresponding clusters need to update the information. Thus, local changes need not be seen and updated by the entire network, and information processed and stored by each mobile node is greatly reduced.

## 1.7 CLUSTERING IN MANET

Clustering technique is not a routing protocol, it is a method that aggregates nodes into groups to make network management easier. Cluster organization of an ad hoc network cannot be achieved offline in fixed infrastructure. The implementation of clustering schemes facilitate the performance of protocols for the Media Access Control(MAC) by improving spatial reuse, throughput, scalability and power consumption.

Clustering provides multiple benefits in MANETs. Firstly, a cluster structure ensures scalability and load balancing in MANETs. It also increases the system capacity by facilitating the spatial reuse of resources. In multi cluster structure, two non overlapping clusters may set up same frequency if they are not adjacent. Also it elects a mobile host with special features known as cluster head for enhanced co-ordination of transmission activities. This mitigates the transmission collision of mobile nodes and hence helps in saving energy and resources. Secondly, it restricts the generation and spreading of routing information by forming a virtual backbone for inter-cluster routing comprising of cluster heads and cluster gateways. Lastly, clustering gives a stable and smaller vision of ad hoc network as when a mobile node changes it's attaching cluster the entire network need not be aware of the local changes in a set of nodes only the mobile nodes in the corresponding cluster reflect the changes. Therefore each node stores and processes a fraction of the total network routing information, thus saving a lot of resources.

The objective of clustering is to maintain a connected cluster. Nodes play different roles in clustering techniques and there are three types of nodes. They are defined as follows:-

1. Ordinary nodes
2. Cluster head
3. Cluster gateways

### 1.7.1 ORDINARY NODES

These are members of a cluster who do not have any neighbouring mobile node belonging to a different cluster.

### 1.7.2 CLUSTER HEAD

These are elected from among the ordinary nodes which form the network backbone and also act as local coordinators to perform power control and routing functions. Although they don't possess any special hardware, they represent dynamic and mobile behaviour. Cluster head exhibits some special features in comparison to other ordinary nodes. Nodes with higher degree of relative stability and greater power backup are elected as clusterheads. The primary task of cluster head is to discover the routes for distant messages and forward inter-cluster packets. Firstly a packet originating from an ordinary node, is directed to its cluster head. If the destination is in the same cluster it is forwarded to the destination node. If it is in a different cluster, then the cluster head of the source node routes the packet within the network to the cluster head of the destination node.

### 1.7.3 CLUSTER GATEWAY

These are ordinary nodes in a non clustered state always located at the periphery of a cluster. Such types of nodes listen to transmissions from other nodes in different clusters. They act as routing devices because they help in transporting the packet from one cluster to another.

## 1.8 COST INVOLVED WITH CLUSTERING

Constructing and maintaining a cluster structure requires additional cost compared to flat structure MANETs. The analysis of cost of clustering scheme is carried out quantitatively or qualitatively to outline the benefits and drawbacks of the clustering technique.
The cost associated with clustering is explained as below:-

 1. In a dynamically changing cluster structure due to frequent change in network topology the information related to clusters vary drastically. The consequent transfer of message packets consumes substantial bandwidth and depletes the energy possessed by mobile nodes. This further disables the upper layer applications which cannot be implemented with handful of resources.

 2. Re-clustering may take place in some clustering schemes due to abrupt local instance such as movement of mobile node to another cluster or death of a mobile node or even shut down of clusterheads, thus leading to re-election of clusterheads. This is known as ripple effect of re-clustering which stimulates this effect of re-clustering over the entire network.

 3. Clustering scheme is divided into two stages: cluster formation and maintenance. The formation stage assumes that the mobile nodes are static. With a frozen period of motion, each mobile node can obtain accurate information from neighbouring nodes, which may not be applicable in real time scenario.

## 1.9 CLUSTER FORMATION

In this clustering scheme, each node can be at most at K hops from the cluster head. However, as long as the node is reachable from the other members of a cluster, it retains its membership. The cluster is formed when a node boots up in the network. After booting up, the node broadcasts a cluster solicitation message to its neighbours. It waits for a timeout period of 't' and if it does not get any reply within that period , it declares itself as cluster head. If it receives a reply message, it examines the hop distance and residual energy fields in the received replies. If the hop distance is less than or equal to K and residual energy is greater than or equal to the threshold value, then the node sends a cluster acceptance message to all the nodes from which replies have been received and merges the cluster with the minimum hop count . If either of these conditions fail, the node declares itself as the cluster head. If the node does not declare itself a cluster head and receives more than one replies to its cluster solicitation message , then the node declares itself a gateway and informs all the nodes from which it received replies. This is because multiple replies to its cluster solicitation message imply that the node is within the transmission range of multiple clusters.

## 1.10 CLUSTER ANALYSIS

It is a technique of dividing data into groups (clusters) which are useful and meaningful. Whether for understanding or utility, cluster analysis played a very crucial role in the field of statistics, biology, information retrieval, machine learning and data mining. Cluster analysis provides an abstraction from individual data objects to the clusters in which those data objects reside.

In mobile ad hoc networks we implement clustering for usefulness and deduction of communication overhead. We avoid declaring clusterheads which are 1 hop away from each other and too many clusterheads in space. To overcome these we opt for well known cluster discovery algorithms which will enable us to designate uniformly distributed clusters over the network. In MANETs clustering technique limit the information flow to local domain. Another aim is to discover neighbouring nodes of the clusterheads with less information overhead which is possible utility of this technique.

There are various approaches to divide objects into sets of clusters and different type of clusters. Basically, there are three most popular algorithms of cluster formation. i.e. K-means, agglomerative hierarchical clustering and DBSCAN.

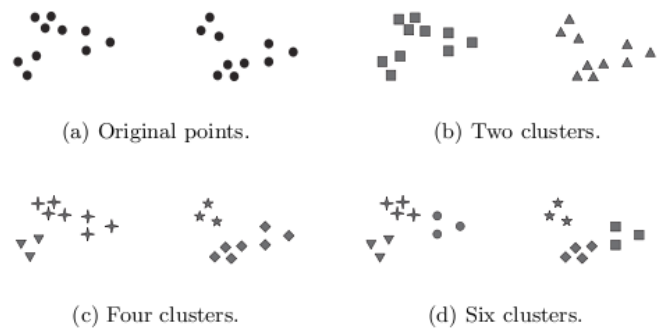

(a) Original points.　　　　(b) Two clusters.

(c) Four clusters.　　　　(d) Six clusters.

**K-MEANS**

This is a prototype-based, partitioning clustering technique that attempts to find a user-specified no. of clusters (k), which are represented by their centroids.

Agglomerative Hierarchical Clustering. This clustering approach refers to a collection of closely related clustering techniques that produce a hierarchical clustering by starting with each point as a singleton cluster and then repeatedly merging the two closest clusters until a single, all encompassing cluster remains. Some of these techniques have a natural interpretation in terms of graph-based clustering, while others have an interpretation in terms of a prototype-based approach.

**DBSCAN**

This is a density-based clustering algorithm that produces a partitional clustering, in which the number of clusters is automatically determined by the algorithm. But this algorithm has a disadvantage that the points in low-density regions are classified as noise and omitted; thus, DBSCAN does not produce a complete clustering.

# 2. RELATED WORK

In this section, we describe some of the most important clustering schemes. Several approaches to cluster formation have been proposed and surveyed. Here we briefly review the salient features of a few major approaches.

## 2.1 DS BASED CLUSTERING

In this scheme, routing is done based on a set of dominating nodes which function as the cluster heads and relay routing information and data packets. The vertices of a Dominating Set (DS) act as cluster heads and each node in a MANET is assigned to one cluster head that dominates it. A DS is called a Connected Dominating Set (CDS) if all the dominating nodes are directly connected to each other. Wu's CDS Algorithm gives details for the formation of CDS. Later, Chen's Weakly Connected Dominating Set (WCDS) algorithm was proposed which relaxed some of the rules of Wu's Algorithm to form a Weakly Connected Dominating Set. There are many disadvantages with the CDS algorithm. The cluster head in CDS algorithm dissipates more power as compared to other nodes in the cluster since all inter-cluster routing and forwarding happen through it alone. Hence it has a shorter lifespan than the other nodes in the cluster. The cluster head re-election is done after the cluster head dies or moves out of the range of the Cluster. This re-clustering incurs a large communication (and power dissipation) overhead.

## 2.2 POWER AWARE CLUSTERING

A CLUSTERPOW algorithm was proposed in which dynamic and implicit clustering is done on the basis of transmit power level. The transmit power level is the power level required to transmit each packet. The transmit power level to a node inside the cluster is less as compared to the level required to send a node outside the cluster. So here the clustering is done keeping the nodes with lower transmit power level together. The primary drawback of their scheme is that there is no cluster head or cluster gateway. Each node here has routing tables corresponding to different transmit power levels. The routing table for a power level $Pi$ in a node is built by communicating with the peer routing table of the same power level at another node. The next hop to route the packet is determined by consulting the lowest power routing table through which the destination is reachable. Thus, this suggests that each node should know the route to other nodes and also know the transmit power level at which a destination node is reachable. This leads to the overhead of collecting the state information and building many routing tables for each power level in a node. There were also other algorithms such as Wu's Algorithm [4] which try to build the DS keeping power as criteria to choose the cluster head. But this scheme also does not overcome the basic drawback of DS based clustering algorithm.

## 2.3 MOBILITY BASED AND WEIGHTED CLUSTERING

Some clustering schemes have been proposed keeping mobility as a metric for cluster construction. In mobility aware clustering, cluster architecture is determined by the mobility behavior of mobile nodes. In such schemes, a cluster is formed by grouping mobile nodes moving with the same velocity. This results in the formation of highly connected intra-cluster links. MOBIC was proposed which takes aggregate local mobility as the metric for cluster formation. Each node broadcasts two hello packets, separated by a time interval, to its neighbors. Every node calculates the relative mobility of each of its neighbors using the signal strength of the hello packets received from each of them.

Each node then calculates its aggregate mobility as the average of the relative mobility of its neighbors and broadcasts it to the other nodes. The node with the lowest aggregate mobility is chosen as the cluster head. This requires larger communication overhead and a higher latency in cluster formation. There also exist other approaches like combined metric based clustering such as On Demand Weighted Clustering Algorithm. This approach calculates a combined weight factor and uses this metric for the cluster formation. These metric based clustering schemes require explicit control messages for cluster formation to exchange the metric information, thus leading to more communication overhead.

A large variety of approaches for ad hoc clustering have been presented, whereby different approaches typically focus on different performance metrics. This thesis presents a comprehensive survey in the following section of recently proposed clustering algorithms, which are classified based on their objectives. This survey provides descriptions of the mechanisms, evaluations of their performance and cost, and discussions of advantages and disadvantages of each clustering scheme.

# 2.4 SURVEY OF CLUSTERING   ALGORITHMS

| Serial No. | Clustering Scheme | Scalability | Power Aware | Mobility/ Scalability Aware | Low Maintenance |
|---|---|---|---|---|---|
| 1. | Lowest Id | No | No | No | No |
| 2. | Highest Degree | No | No | No | No |
| 3. | Max-Min d - cluster [ d is the maximum no. of hops that a node can be away from the cluster head] | No | No | No | No |
| 4. | K-hop connectivity [All the nodes of a cluster are at most k-hops from CH.] | No | No | No | No |
| 5. | Adaptive cluster load balancing | Yes | No | No | No |
| 6. | Adaptive multi-hop | Yes | No | No | No |
| 7. | Mobility based D-hop clustering (Mob hop) | Yes | No | Yes | Yes |
| 8. | Mobility based Metric | No | No | Yes | Yes |
| 9. | Mobility based framework | No | No | Yes | No |
| 10. | Least cluster change(LCC) | No | No | No | Yes |
| 11. | Load balancing (LBC) | No | Yes | No | No |
| 12. | Power aware connected dominant set | No | Yes | No | No |
| 13. | Clustering for energy conservation | No | Yes | No | No |
| 14. | Weighted Clustering algorithm(WCA) | Yes | Yes | Yes | No |
| 15. | Entropy-based weighted clustering algorithm | Yes | Yes | Yes | Yes |
| 16. | Weight-based adaptive clustering(WBACA) | No | Yes | Yes | Yes |

| Serial No. | Clustering Scheme | Cluster head Selection Criteria | Benefits | Drawbacks |
|---|---|---|---|---|
| 1. | Lowest Id | Node with minimum ID which is distinct. | Simple to implement. | 1> Certain nodes are prone to power drainage due to serving as CH for longer period.<br><br>2> Generates more CH than necessary. |
| 2. | Highest Degree | Node with Highest connectivity value among its direct neighbours. | Less hop to fulfil a request. | 1> Due to high mobility frequent change of topology occurs. Hence Congestion in Cluster head and erratic exchange of CH occurs.<br><br>2> Numerous ties between nodes.<br><br>3> Only one hop connectivity.<br><br>4> One cluster can be exhausted serving too many hosts. |
| 3. | Max-Min d-cluster [d is the maximum no. of hops that a node can be away from the cluster head] | 1> Node participates in CH selection based on their node ID.<br><br>2> Two data structures are used ( arrays) ->Winner and Sender for a particular round. There are 2d rounds of flooding.<br><br>3> Once the CH is selected sender node determines the shortest | 1> Less no. of cluster heads (flexible) and hence less traffic.<br><br>2> Operates asynchronously.<br><br>3> Maximizes no. of gateway nodes.<br><br>4> No. of messages sent from each node is limited to a multiple of d rather than 'n' | |

| | | path. | ( total no. of nodes). | |
|---|---|---|---|---|
| 4. | K-hop connectivity [All the nodes of a cluster are at most k-hops from CH.] | 1> One round of flooding (clustering request to all nodes).<br><br>2> Node with higher connectivity is chosen as CH, in case of tie ID is considered to select CH. Each node has 2 tuples (d,ID)<br>d: degree | 1> Obtain minimum no. of clusters and smaller size of dominating set.<br><br>2> K-hop connectivity. | |
| 5. | Adaptive cluster load balancing | 1> Hello Message format is used which has an item 'options'.<br><br>a> If sender node is CH it will assign 'options' to no. of dominating members else it will be reset to 0.<br><br>b> If hello message of CH shows its dominating set is greater than threshold (max. no. one CH can manage) no new node will participate in that cluster. | 1> CH bottleneck phenomenon is eliminated and cluster structure is optimized.<br><br>2> Load balance between various clusters is implemented.<br><br>3> Resource consumption and information transmission is distributed uniformly among all clusters. | |
| 6. | Adaptive multi-hop | Upper and lower bound (UB & LB) are maintained which specify maximum and minimum no. of nodes in a cluster (based on network size and mobility etc.)<br><br>1> If no. of nodes in a cluster <UB and >LB then | 1> Load balancing algorithm.<br><br>2> Prevents formation of too many clusters<br><br>3> Gateway communicates with other neighbouring gateways in different | 1> Does not address how to reward a node as CH for a newly formed cluster. |

| | | merge occurs.<br><br>2> If no. of nodes > UB then divide of cluster occurs.<br><br><br>3> While flooding own ID, CH ID, status (type of node) are broadcasted to other nodes within the same cluster.<br>By this every node contains cluster information. | clusters and reports to its CH. Thus the CH is acknowledged about the no. of mobile nodes of each neighbouring clusters. | |
|---|---|---|---|---|
| 7. | Mobility based D-hop clustering (Mob hop) | 1> Based on the received signal strength by each node.<br><br>2> Cluster formation is mostly by message driven.<br><br>3> Lowest value of local stability among neighbours.<br><br>4> Divides network into d-hop cluster.<br><br>This algorithm is based on mobility metric and the diameter of a cluster is adoptable with respect to node mobility.<br>This algorithm is conducted in three phases (Discovery, merging and cluster maintenance). | 1> Focuses to make cluster diameter more flexible.<br><br>2> It considers cluster maintenance during its third phase. | 1> It assumes each node can measure its received signal strength depending<br>upon the closeness between two nodes. |

| 8. | Mobility based Metric | 1> Mobile nodes having low speed relative to their neighbours have higher chances to be CH.<br><br>2> By calculating variance of mobile nodes speed relative to each of its neighbours and the aggregate speed of a local node is measure.<br><br>3> A node with minimum aggregate value is chosen as CH. | 1> Timer is used to avoid Frequent reclustering. | 1> In cluster maintenance speed criteria is ignored.<br><br>2> This scheme is effective only for high way situation when neighbours move constantly but poor performance if random movement is present. |
|---|---|---|---|---|
| 9. | Mobility based framework | 1> Clustering is based on (a,t) criteria.<br><br>Note: 'a' is the probability that each mobile node in a cluster has a path to every other node that will be available over some time period 't' regardless of hop distance. | 1> Adaptable to high rate of topology changes.<br><br>2> Achieved by prediction of future state of network links.<br><br>3> A metric captures dynamics of node mobility makes the scheme adaptive with respect to node mobility.<br>This scheme supports hybrid routing strategy hence it is more responsive and effective if mobility is less and efficient if mobility is high. | |

| 10. | Least cluster change(LCC) | 1> Nodes with lowest ID are chosen as CH.<br><br>2> Reclustering is done if and only if :<br><br>a>Two nodes move close to each other and one gives up<br>.<br>b>A mobile node cannot access any CH. Hence reclustering occurs based on LIC. | 1> It doesn't recluster the nodes from time to time in order to satisfy specific characteristics of clusterheads.<br><br>2>It doesn't periodically check that if a new node of higher degree is introduced. | |
| --- | --- | --- | --- | --- |
| 11. | Load balancing (LBC) | 1> Each mobile node has a variable Virtual ID (VID). Initially the VID equals to ID of the node.<br><br>2> Mobile nodes with highest ID in the local area win the CH role.<br><br>3> When two CH move into the reach range the one with the higher VID wins the CH role.<br><br>4> When a CH resigns a non CH with a largest VID in the neighbourhood resumes the role of CH whose previous cluster head serving time is the shortest in the neighbourhood. | 1> This algorithm limits the maximum time limit that a node can serve as a CH continuously so when a CH exhausts its duration value it resets its VID equal to 0 and becomes a non CH. | 1> Cluster head serving time alone may not be a good indicator of energy consumption of a mobile node. |

| 12. | Power aware connected dominant set | 1> Energy level instead of ID or node degree is used to determine whether a node should serve as a cluster head.<br><br>2>A mobile node can be deleted from the DS when its close neighbour set is covered by one or more Dominating neighbours and at the same time it has less residual energy than the neighbours. | 1> Energy efficient clustering scheme which decreases the size of dominating set without impairing its functions.<br><br>2> The unnecessary mobile nodes are excluded from the dominating sets saving their energy consumed for serving as CHs, as mobile nodes inside a DS consumes more power than outsize the DS. | 1> It cannot balance great difference of energy consumption between dominating nodes and non-dominating nodes because its objective is to minimize the dominating set rather than balancing the energy consumption among all mobile nodes.<br><br>2> Mobile nodes in the DS still likely deplete their energy at a much faster rate. |
| --- | --- | --- | --- | --- |
| 13. | Clustering for energy conservation | 1> Master: Forms a cluster with connections to slaves and can serve only limited no. of slaves.<br>Slave: No direct connection among them.<br><br>2> Two schemes are present, single-phase and double-phase clustering.<br><br>3> In single phase :<br>a>master node pages the slave node with maximum energy.<br>b>Slaves send ACK of signals.<br>c>Slaves receiving signals from only one master are allocated channels if free | 1> Minimize the transmission energy consumed by all (m-s) pair.<br><br>2> Serve as many slaves as possible to sustain for longer lifetime resulting to better performance. | 1> Paging process before each communication round is expensive with energy.<br><br>2> The criteria for selection of master node is not specified. |
| | | channels remain and other slaves are allocated with decreasing power signals.<br>d>Slaves try to communicate with nearest master.<br>4>Slaves which don't get channel are ignored and they are looked after in double-phase clustering based on the received signal (second round) strength. | | |
| 14. | Weighted Clustering algorithm(WCA) | 1> Invoked when current DS is unable to cover all nodes.<br>(A threshold value limits max. no. of cluster nodes)<br><br>$W_V=W_1Del_V+W_2D_V+W_3M_V+W_4P_V$<br>$W_1+W_2+W_3+W_4=1$<br>Where,<br>• $Del_V$ =[ dv-delta]<br>• dv=degree of node<br>• delta=predefined threshold<br>• $D_V$=sum of distance from all its neighbours<br>• $M_V$=mobility(avg. Speed during time T)<br>• $P_V$=how much battery has been consumed.<br><br>2> The node with minimum $W_V$ is declared as CH. | 1> Algorithm is not periodic and therefore avoids communication overhead.<br><br>2>Considers many factors at once. | 1> In highly mobile scenario there is high frequency of reaffiliation. Hence, network overhead increases.<br><br>2> More frequent recalculation of value. Hence, communication overhead increases.<br><br>3> Clusterheads may not be separated over the network as they can be 1-hop neighbours. |

| 15. | Entropy-based weighted clustering algorithm | 1> Overcomes disadvantages of WCA as it forms a stable network. 2>Uses an entropy based model for evaluating route stability and cluster head election. | 1> It defines better indication of stability and mobility of ad hoc network as it evaluates entropy. | |
|---|---|---|---|---|
| 16. | Weight-based adaptive clustering(WB ACA) | 1> It considers :<br>• M:Mobility of the node<br>• B:Battery power<br>• $T_X$:Transmission Power<br>• D:Degree Difference<br>• $T_R$: Transmission rate<br><br>$W_N=W_1*M+W_2*B+W_3*T_X+W_4*D+W_5/T_R$<br><br>2> Node with smallest weight is chosen as CH. | 1> All the ordinary nodes are one hop away from cluster head.<br>Overlapping clusters are connected using gateway nodes. | |

# 3. PROBLEM FORMULATION

## 3.1 PROBLEM DESCRIPTION

In MANET there is no fixed infrastructure therefore the mobile hosts communicate over multihop wireless links. It is based on dynamic topology as nodes are connected dynamically in an arbitrary manner. This gives rise to a problem as the wireless links of the network vary periodically and have significantly lower capacity than the wired links.

Moreover in MANETs, mobile nodes rely on battery power and storage capacity which are limited resources. In large networks MANET faces scalability and mobility issues which lead to high message passing and hence leads to the possibility of packet losses in the network.

## 3.2 VISUALIZATION OF PROBLEM

To overcome the above stated problems in MANET, an energy aware, scalable , mobility based clustering algorithm is developed which is independent of the routing protocol.

### 3.3 SOFTWARES USED FOR IMPLEMENTATION

| Name | Utility |
|------|---------|
| NS 2.35-RC7 | Open source network simulator |
| GNU PLOT | To plot the corresponding observations |
| Clustering Framework | C++ source modified to implement our clustering algorithm. |
| Linux ( UBUNTU) | Open source operating system used to run all the softwares. |

## 3.4 VISUALIZATION OF PROBLEM using NS2

NS2 is widely used open source network simulation software. We implement clustering in MANET which is independent of the routing protocol, with the help of *The Clustering Framework* (discussed in the following section). The algorithm works perfectly with AODV, DSDV and other protocols. The available network topology is supplied to the clustering algorithm which subdivides the network into clusters which improves the message passing and throughput.

## 3.5 THE CLUSTERING FRAMEWORK

A clustering framework has been used to implement the clustering algorithm. Clustering framework is a library that can used for developing clustering algorithms. A clustering algorithm is a layer placed between the MAC and Link layers in the mobile node structure of ns2.

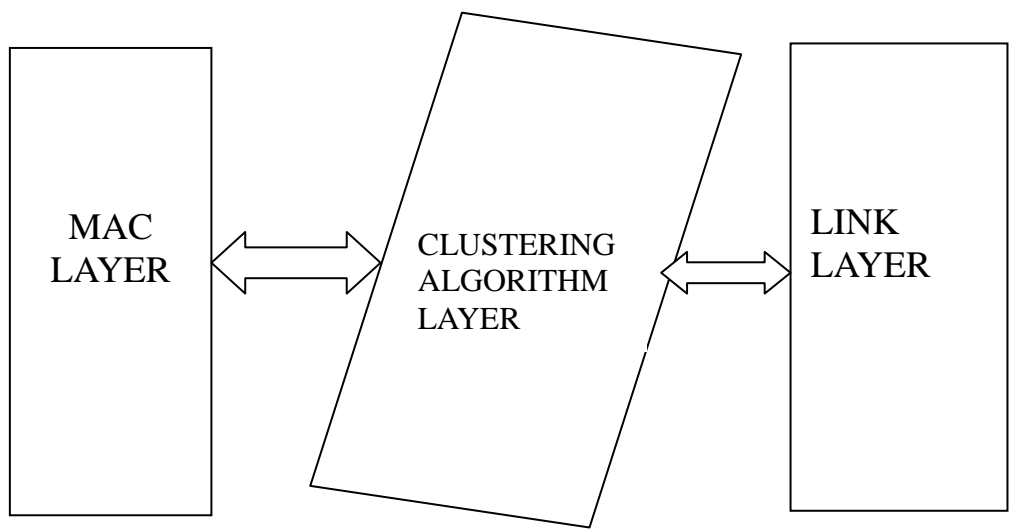

**FIG : LAYER  REPRESENTATION IN NETWORK MODEL**

The aim of a clustering algorithm is to create an overlay over the network visibility graph of a given topology. A network visibility graph is the graph obtained by having a link between two nodes that can communicate bi-directionally.

The MAC layer provides primitives to send messages either as a unicast message from one node to another specific node or as a broadcast message from one node to all its neighbouring nodes. A clustering algorithm require nodes to exchange messages to create a rich backbone and then try to reduce the number of backbone nodes by exploiting some topological information. Therefore the final result of the clustering algorithm is given by the sequential application of two distributed algorithms. The first distributed algorithm is able to find a first backbone and second one uses the information collected to improve the choice.

For the simplified analysis of the algorithms, a C++ class called separator is responsible to run a distributed algorithm , wait for notification from each node and run the clustering algorithm after  the reception of the notification by every node in the network. Each distributed algorithm of a clustering algorithm is called Clustering Sub Layer (CSL).

The separator class is responsible to run a distributed algorithm which waits for termination notification and run the following algorithm after the reception of the notification by every node in the network. It is just a simulator helper that allows coordinating difference simulation steps.

### 3.5.1 FRAMEWORK  DISTRIBUTED ALGORITHM

The Clustering Sub Layer (CSL) of a clustering algorithm is simulated as follows:

- When simulation starts the first CSL of each node in the network receive a start-Module message (there are no assumptions on the ordering of execution). The start procedure launches the clustering procedure simultaneously at each node (where the execution time of each algorithm is the same simulation time for each node).

- Each CSL exchanges messages with neighbors using primitives provided by the MAC layer (distributed algorithm implementation).

- When a CSL installed in a node terminates its procedure (when for example the node contacted every neighbor to exchange some useful information) it calls the method endModule.

- All information collected by the dumper of the current clustering layer are dumped on the screen.

- If the current algorithm was the last one, the simulation end and the global statistics collected are printed out, otherwise the simulator restarts the execution from step (1) by using the next CSL that defines the clustering algorithm.

- Each CSL can dump its own results on standard output as well as on the global result collector.

# 4. ENERGY AWARE, SCALABLE, K-HOP BASED CLUSTER FORMATION ALGORITHM

## 4.1 ALGORITHM

The algorithm takes care of a proper and optimized cluster head selection algorithm considering factors such as :

- battery power
- scalability issues
- number of message passing
- random topology leading to high mobility
- bandwidth of its wireless   capabilities
- processing power
- memory capacity and density of nodes (i.e. load balancing).

### 4.1.1 ASSUMPTIONS:

- Every node is capable enough to compute its priority.
- Factors such as battery power, mobility, memory capability and processing power do not vary rapidly with respect to time.
- There are no malicious/byzantine faults possible in the system.
- The system continues to be active state in the network until and unless the battery power is exhausted. After that the node dies.

### 4.1.2 PSEUDOCODE

**Step 1:** All the nodes compute their priority P based on the following functions.
This calculation is repeated in background in a definite timestamp(T1).

Priority P is calculated as follows:
$P = W1 * B + W2 * 1/M + W3 * MC + W4 * NS + W5 * PP$

W1, W2, W3, W4, W5 -> are the integral weights of contribution to the priority .

| | | |
|---|---|---|
| B | -> | Battery power |
| M | -> | Mobility |
| MC | -> | Memory Capacity |
| NS | -> | Number of Single hop nodes |
| PP | -> | Processing Power |

The number of single hops must be $<=$ threshold number. If it is greater NS will hold negative value.

Every node maintains an integer in its stable memory stating the time for which it has served.

**Step 2:** If any node at any point of time sense that there is no cluster head (CH) available, it starts an "elect me" cycle. It sends a message M(E, P, serve time) to all other nodes. The message M has a format as follows:

Message= M ("Unique Message",P, serve time).

The parameter "Unique Message" defines the state of the algorithm. i.e. 'E' for elect me,'IAMC' for declaring I am the cluster head, 'Q' when a cluster head quits and 'S' to stop elect me cycle.

The parameter P is computed priority at the instant of sending message.

The parameter "serve time marker" implies the duration that the node or site has actually served as cluster head until it sends the message M.

**Step 3:** The message sending node waits for a timeout period of T2,

if (it receives any reply from a higher priority and lower serve time node within T2), then
then it withdraws from the "elect me" cycle and continues behaving as a normal node and waits for a time out period of T3 . If this node does not receive an "elect me" message from that higher priority node within T3 then it again starts its own elect me cycle.

Else if (it receive any reply from a node with higher priority and higher serve time within T2) then
it declares itself as the cluster head sending a message M(IAMC, P, serve time).

Else /* No reply has been received*/
it declares itself as the cluster head sending a message M(IAMC, P, serve time).

When a node receives a message M(E, P, serve time) from a node which initiated its elect me cycle then the receiving node compares its priority with P.

If (priority of the receiving node >=P)
then it sends a reply to the sending node in the form M(S,P1,serve time1) where P1 and serve time1 are the priority and serve time of the receiving node respectively.

Else it does not send any reply.

**Step 4:** The higher priority node repeats the same action as in step 3.

**Step 5:** If a cluster head has served more than a time-slot(T4 or it's battery power is <10% whichever is minimum) time then it sends a message M(Q,0,serve time ) and sets its priority 0 for a time out period (T5)Go to step 2.

Note: If a node receives two or more than two "Unique message" => IAMC then it is designated as gateway node.

### 4.1.3 ADVANTAGES OF THE ALGORITHM:

1. Fault tolerence: All the nodes do not need to take part in an cluster head election algorithm. Even if a node dies, system continues with its activities.
2. Scalability: A threshold value limiting the number of single hop nodes from a nominee of clusterheads is considered to limit number of nodes in a cluster.
3. Power Efficient: A timeout time T4 is considered so that a node is not exhausted with all it battery power in serving as a cluster head node.
4. Low maintenance: Even after a clusterheads dies, a new cluster head will be selected as soon as a node senses that there is no cluster head in its cluster.
5. Low message passing: Hence traffic overhead is less because only higher priority nodes reply.
6. Availability: There cannot be any cluster without a cluster head for a timeout period .

### 4.1.4 DISADVANTAGES OF THE ALGORITHM:

1. It is assumed that the mobility of the nodes is less; therefore if the nodes are highly mobile then this algorithm does not perform efficiently due to erratic exchange of clusterheads.
2. It is assumed that node can perfectly find out nodes which are at single hop distance with precision.

### 4.1.5 EXAMPLE OF THE ALGORITHM

P= W1 * B+ W2 * 1/M + W3 * MC + W4 * NS + W5 * PP
Assuming, W1=3, W2=4, W3=2, W4=2, W5=1

| Node | Battery Power(B) | Mobility(M)change in pos/time | Memory Capacity(MC) [In scale of 10] | Single hop nodes(NS) | Processing power(PP)[in scale of 10] | Priority(p) | Serve Time (in mins.) |
|------|------|------|------|------|------|------|------|
| 1 | 60%=0.6 | 6 | 3 | 2 | 10 | 32.46 | 5 |
| 2 | 80%=0.8 | 4 | 4 | 1 | 5 | 18.4 | 7 |
| 3 | 25%=0.25 | 1 | 10 | 1 | 7 | 33.75 | 12 |

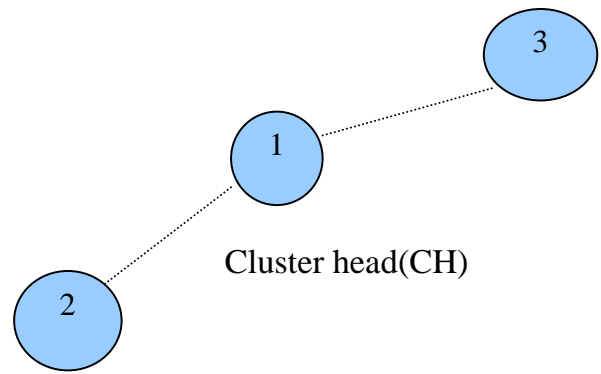

Cluster head(CH)

## 4.1.6 THEORITICAL RESULTS OF THE ALGORITHM

1. Initially there is no cluster head in the system.
2. All the nodes calculates their priority in t=0 and keep updating the priority in every (T1=1 minute)
3. Node 3 senses that there is no cluster head in the network. It starts an elect me cycle and sends a message M(E,33.75,12) to node 1 and node 2(via node 1).Since it has the highest priority it declares itself as a cluster head for a timeout period T4 or until its battery power is <10% .
4. It sets its priority as 0 for a time out time T5 by sending a message M(Q,0,12+T4) .
5. Again a new CH selection algorithm is executed by Node 2.Since node 2 has lower priority it receives a reply from node 1 and terminates its elect me cycle.
6. Now Node 1 runs an elect me cycle and designates itself as a cluster head and runs for T4 time.
7. This continues.

# 5. SIMULATED RESULTS AND PERFORMANCE EVALUATION

## 5.1 EXPERIMENT DUMP FORMAT FOR SIMULATION RESULTS

*Sample dumped data generated by the simulated code :*

### 5.1.1. COMMAND
*$ns Desktop/template/test_program.tcl -t Desktop/topologies/ -N 100 -I 4 -d -k 1*

The command is explained below in the following table:

| |
|---|
| *# test_program.tcl -> The file which is being simulated*<br>*# Desktop/topologies/ -> is the topology directory*<br>*# -N 100 -> Defines the number of nodes*<br>*# -I 4 -> Index of the file*<br>*# -d -> Degree is being considered*<br>*# -k -> k value is 1* |

### 5.1.2. CONTENTS OF THE DUMPED FILE

*# The following entry is a randomly generated topology file which contains X,Y,Z co-ordinates with or without the destinations of the nodes with speed.*
coordS12N100

BACKBONE
*#Number of nodes*
100

*#The energy data obtained is as follows:*
*#Each row contains 11 values which are explained below from left to right*
*# For the 1st row of data generated, each value is explained in the following table .*

| Value from left | Attribute |
|---|---|
| 0 | Node ID |
| 0 | Unicast Bytes Transmitted. |
| 0 | Unicast Messages Transmitted. |
| 32 | Broadcast Bytes Transmitted. |
| 3 | Broadcast Messages Transmitted. |
| 0 | Unicast Bytes Received. |
| 0 | Unicast Messages Received. |
| 36 | Broadcast Bytes Received. |
| 3 | Broadcast Messages Received. |
| 0.626062 | Execution Time. |

| 7.30846e-05 | Consumed Energy. |
|---|---|

**The data generated by the experiment for every node. Each row represents a node in the topology and each row has eleven values which are explained in the table above:**

```
0 0 0 32 3 0 0 36 3 0.626062 7.30846e-05
1 0 0 52 3 0 0 320 18 1.16876 0.000217865
2 104 9 112 4 84 10 656 30 1.64046 0.000925109
3 0 0 56 3 0 0 368 21 0.947235 0.000240857
4 0 0 52 3 0 0 324 18 1.11822 0.000217567
5 0 0 44 3 0 0 188 12 1.01035 0.000158715
6 16 2 60 3 24 2 484 25 1.70145 0.000572205
7 32 3 60 3 24 3 516 25 1.57054 0.000680155
8 0 0 44 3 0 0 220 12 1.10855 0.000163871
9 0 0 48 3 0 0 244 15 0.869515 0.000183839
10 8 1 56 3 12 1 464 22 1.47813 0.000389468
11 0 0 60 3 0 0 520 24 1.1006 0.000281442
12 8 1 64 3 12 1 644 28 1.67767 0.000817716
13 8 1 68 3 12 1 588 30 1.54227 0.000539115
14 0 0 56 3 0 0 444 21 1.1006 0.000251904
15 44 4 48 3 36 4 264 15 1.49886 0.000331164
16 0 0 36 3 0 0 88 6 1.18495 0.000108244
17 8 1 52 3 0 0 360 19 1.72828 0.000461413
18 0 0 68 3 0 0 588 30 1.1006 0.00033085
19 0 0 44 3 0 0 208 12 1.01035 0.000161017
20 0 0 60 3 0 0 472 24 1.58789 0.000500038
21 0 0 48 3 0 0 256 15 1.01035 0.000187332
22 8 1 60 3 12 1 536 25 1.57438 0.000532998
23 0 0 52 3 0 0 388 18 1.00668 0.000223259
24 0 0 44 3 0 0 256 12 1.1006 0.000167895
25 8 1 64 3 12 1 536 27 1.53785 0.000469003
26 24 2 56 3 16 2 424 21 1.58789 0.000502906
27 20 2 52 3 20 2 404 19 1.72828 0.000717296
28 0 0 40 3 0 0 148 9 1.40433 0.00018204
29 16 2 68 3 12 1 668 30 2.51189 0.00114225
30 64 6 64 3 56 6 492 25 2.16257 0.000793486
31 0 0 40 3 0 0 128 9 1.12887 0.000132797
32 0 0 40 3 0 0 148 9 0.869515 0.000131208
33 0 0 48 3 0 0 252 15 1.01035 0.000186872
34 0 0 56 3 0 0 348 21 0.893817 0.000237755
35 8 1 64 3 12 1 564 28 1.49335 0.000516365
36 0 0 64 3 0 0 508 27 1.1006 0.000300852
37 0 0 40 3 0 0 136 9 0.759089 0.000114739
38 0 0 56 3 0 0 356 21 0.947235 0.000239476
.....................................................................
97 0 0 44 3 0 0 216 12 1.09875 0.000149832
98 20 2 88 4 28 3 420 21 1.72828 0.000769019
```

```
99 8 1 68 3 12 1 692 31 1.67767 0.000687766
```

*#Following row of data has two values:*
*1> The first value is the total number of packets transmitted by physical layer which considers data retransmissions for the unicast messages. ( 381 as in the example below)*
*2> The second value implies the total amount of bytes transmitted by physical layer respectively. (28266 as in the example below)*
381 28266

*#Following value implies the types of colours assigned to the nodes*
  2
*# **Black** is the colour given to the Cluster head and **white** is the colour given to a normal node (In some simulations colour=Grey).*

```
black 1 3 4 7 13 22 30 34 35 38 40 42 43 45 48 49 51 57 58 60 62 64 67 71 73 74 77 78 79 81 83
84 86 87 89 90 92 93 96 99
white 0 2 5 6 8 9 10 11 12 14 15 16 17 18 19 20 21 23 24 25 26 27 28 29 31 32 33 36 37 39 41
44 46 47 50 52 53 54 55 56 59 61 63 65 66 68 69 70 72 75 76 80 82 85 88 91 94 95 97 98
```

#Nodes and their connectivity.

*Total number of nodes*
100

**For a sample of data generated experimentally, the node and its connectivity is explained below:**

| Entry | Node | Connectivity |
|-------|------|--------------|
| 1 64 71 92 | 1 | 1->64,1->71,1->92 (3 wireless links) |
| 0 71 | 0 | 0-> 71 (1 wireless link from 0th to 71st node) |

**Following is the dumped data each row represents a node sorted in ascending order:-**

```
0 71
1 64 71 92
2 99
3 4 34 38 67 73 77
4 3 38 74 77 83
5 86
6 92
7 58 92
8 84
9 38
10 99
11 96
12 99
13 30 40 62
14 99
15 42
16 49
17 92
18 96
19 73
20 30
21 86
22 35 62 99
23 96
24 90
25 40
26 42
27 92
28 42
29 89
30 13 20 42
31 60
32 34
33 73
34 3 32 38 73 77
35 22 81 87 99
36 90
37 86
38 3 4 9 34 77 82
39 57
40 13 25 62 87
......................................................
97 64
98 92
```

99 2 10 12 14 22 35 43 78 89 91

## 5.2 PERFORMANCE EVALUATION

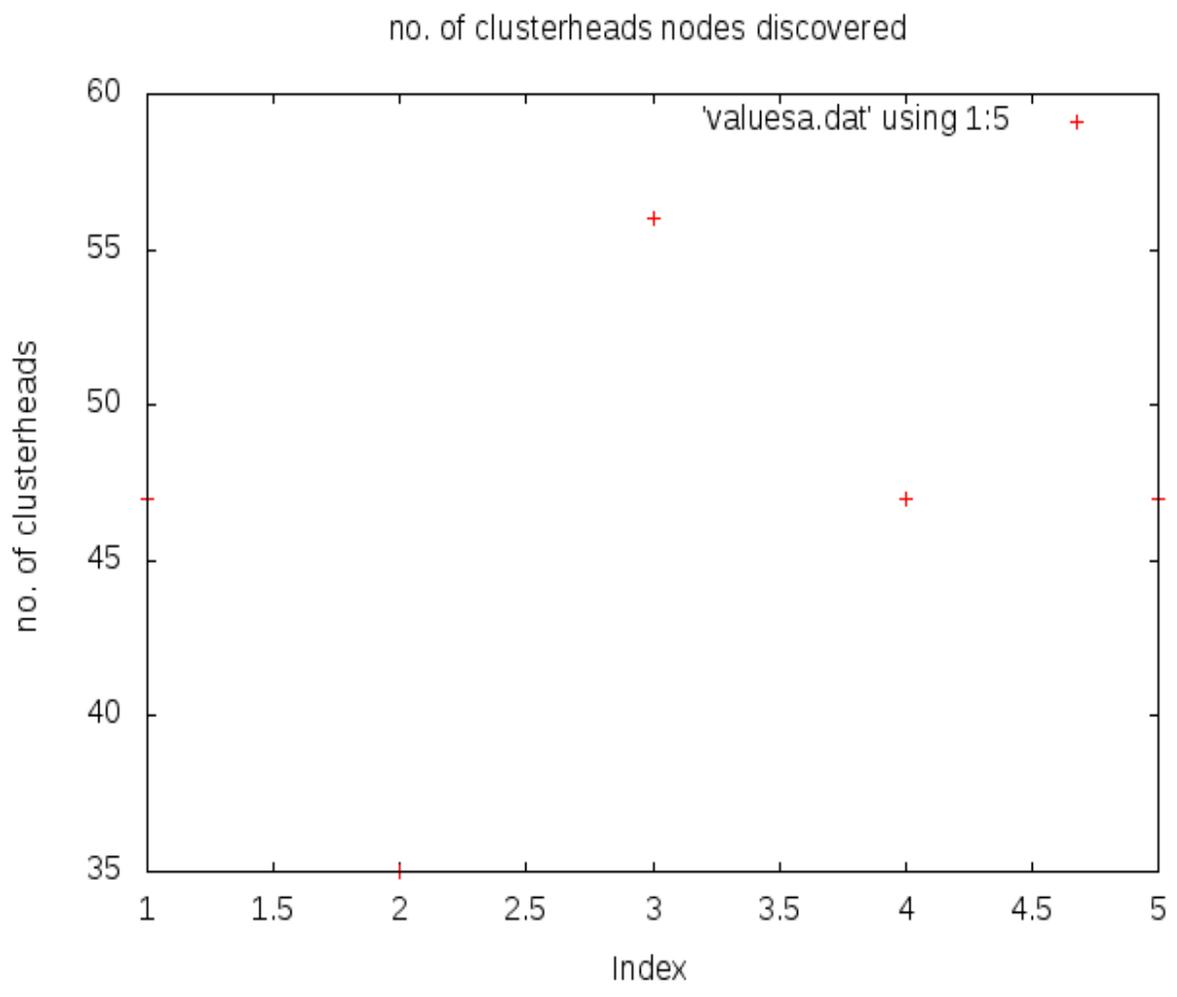

**Fig 5.2.1: Represents the simulation results for 5 different randomly generated topologies. Each topology consists of X, Y and Z co-ordinates of the nodes and N (no. Of nodes) =100 in the above scenario.**

Explanation:
With different topologies where initial position varies with Index but constant N value, number of clusterheads nodes varies from 48 to 58 depending of concentration of nodes .We have considered K=3 in above simulation. This means it follows K hop clustering with that particular K value. Nodes up to 3 hops are connected to the clusterheads.

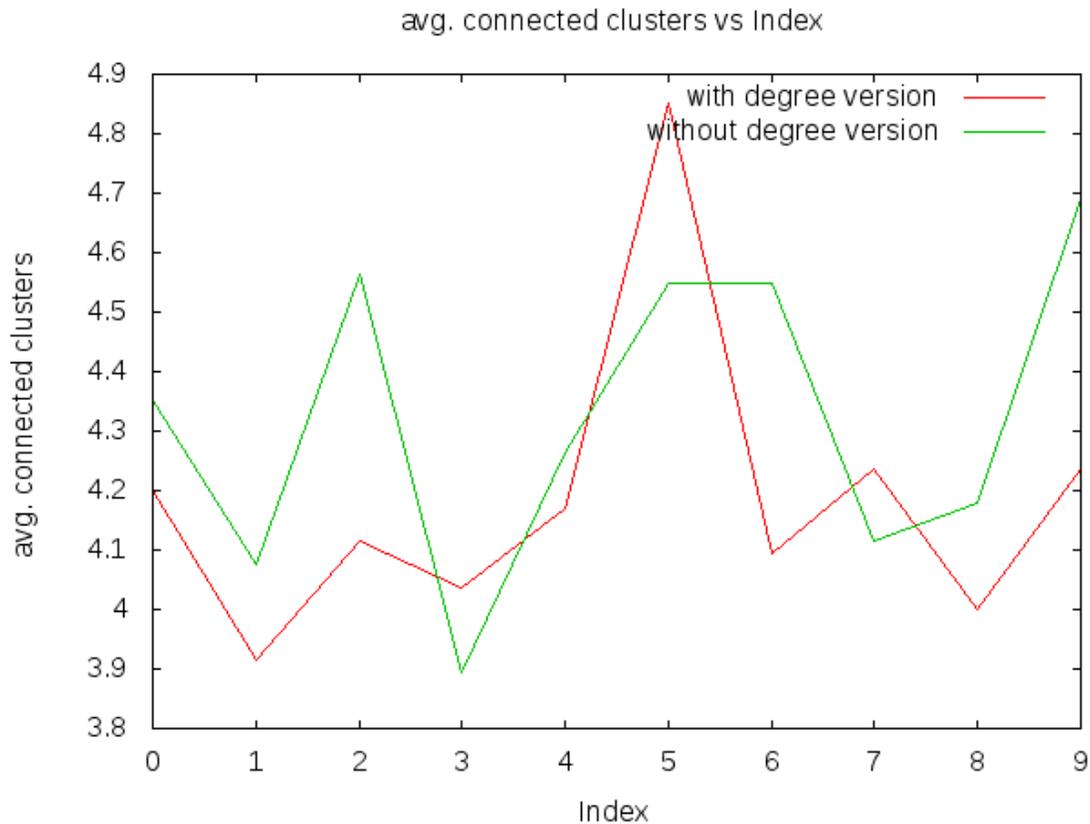

**Fig 5.2.2: Represents the simulation results for 10 different randomly generated topologies. Each topology consists of X, Y and Z coordinates of the nodes and N=100 in the above scenario. Terminologies:**

*Avg. Connected Clusterheads*=connectedClusterHeads/(summation of connectivity of all the clusters)

*Degree version:* It's a special version of our algorithm where degree is considered as one of the criteria for clustering along with the value of K and energy.

Explanation:

With different topologies where initial position varies with Index but constant N value, the general trend in the above simulation denotes that avg. connected clusterheads value is lesser in degree version as compared to non-degree version with few exceptions. This denotes degree version improves the clustering and the average connected

Clusterheads increases and hence the connectivity. But it's dangerous if degree is too high, where congestion at clusterheads will harm the performance of the network and of the node too.

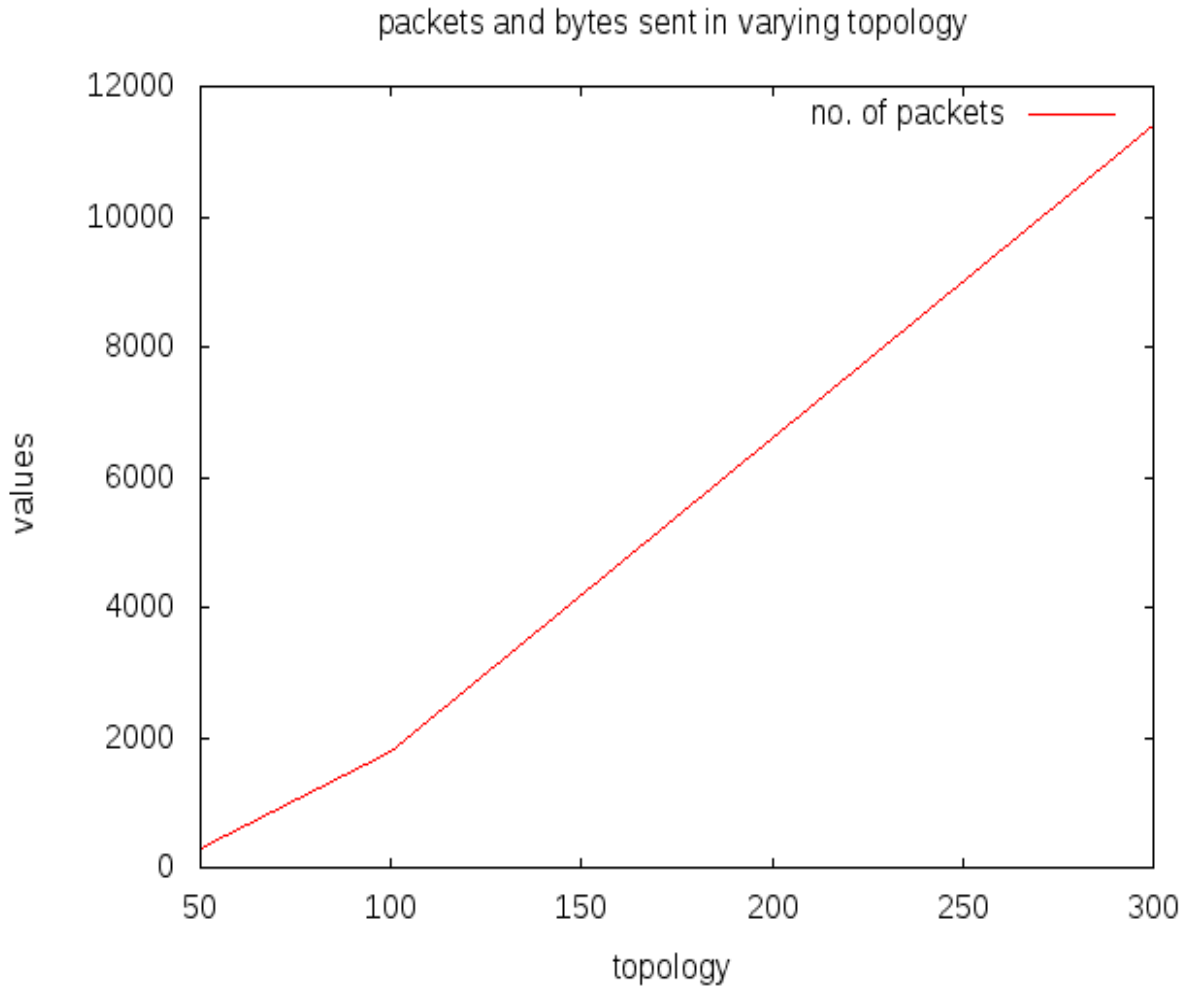

**Fig 5.2.3: Represents the simulation results for 3 different randomly generated topologies of different sizes. Each topology consists of X, Y and Z co-ordinates of the nodes and N (no. of nodes) is 50,100 and 300 in the above scenario.**

Terminologies:
*Packets:* A packet is one unit of binary data capable of being routed through a computer network.
Explanation:
With different topologies where initial position varies with variable N value, as we increase the number of nodes in a topology, number of packets transferred also increases. The graph is a linear instead of an exponential one, this denotes the algorithm is scalable on the basis of data. There is not a very steep rise in number of packets as the number of node increases.

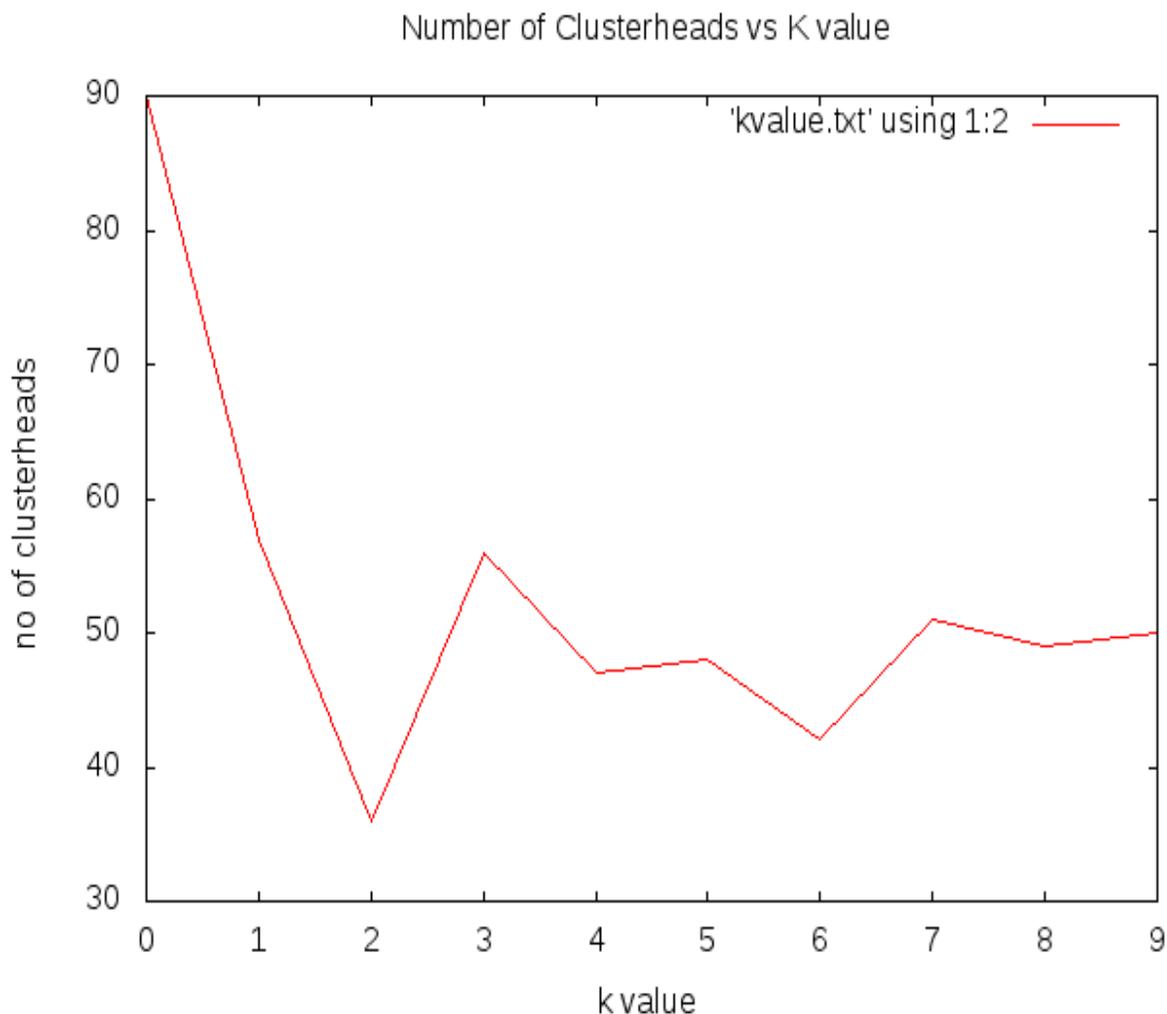

**Fig 5.2.4: Represents the simulation results for 10 different randomly generated topologies with N=100 and varying K value. Each topology consists of X,Y and Z co-ordinates of the nodes and N=100.**

Terminologies:

*K value:* This defines that a node which is at most K hops away from the clusterheads can be connected to the Cluster.

Explanation:

With different topologies where initial position varies with variable N value, As we increase the K value in a topology, number of Clusterheads decreases sharply from 0-2 value of K in a 100 nodes network and with further increase of the variable parameter there is not much change in the number of clusterheads. This denotes that K value should be optimum and different network can result to a better structure with a suitable K values which must be simulated before for better performance.

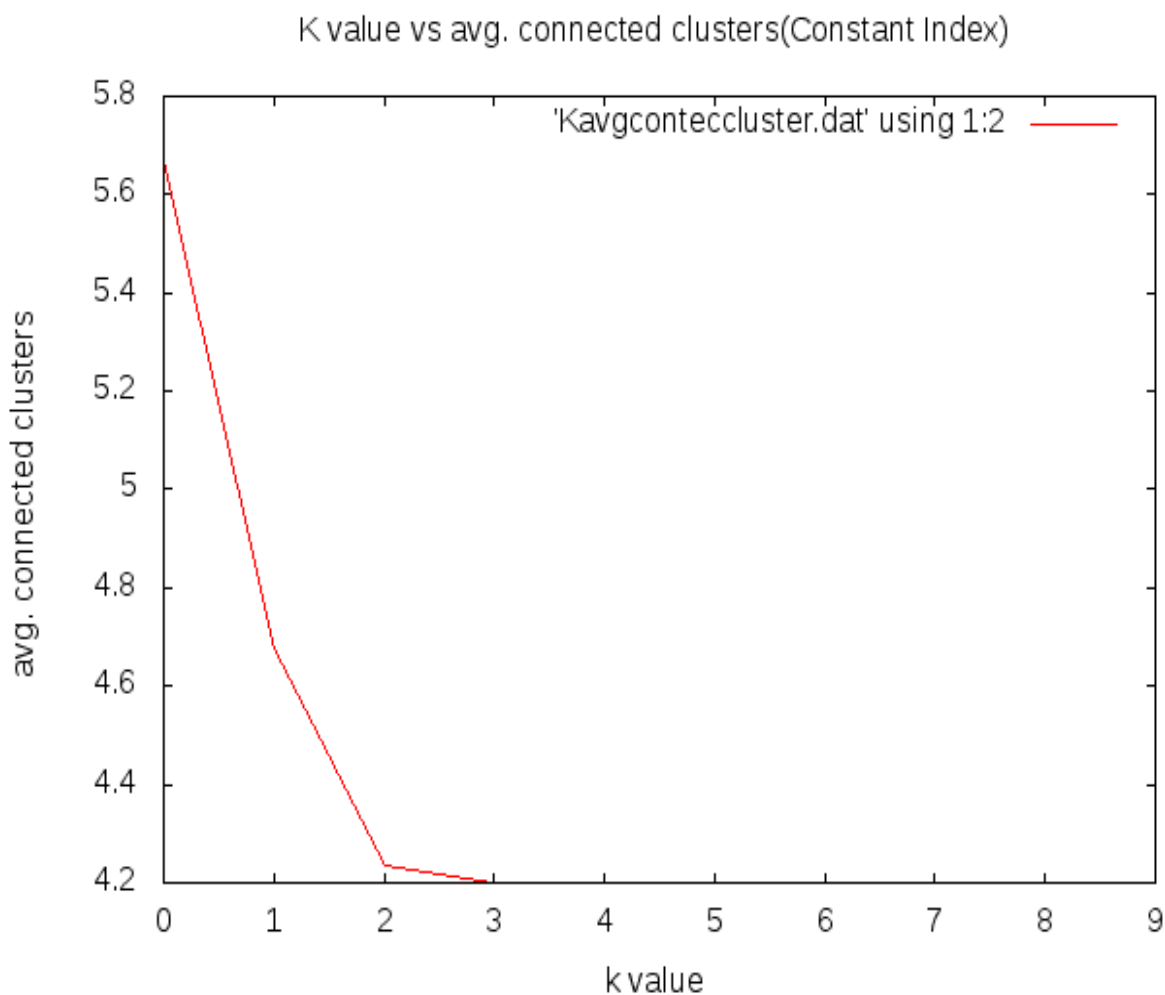

**Fig 5.2.5: Represents the simulation results for 10 different randomly generated topologies with N=100 and varying K value. Each topology consists of X, Y and Z co-ordinates of the nodes and N(no. Of nodes) =100.**

Terminologies:

*K value:* This defines that a node which is at most K hops away from the clusterheads can be connected to the Cluster.

Explanation:

With different topologies where initial position varies with variable N value, As we increase the K value in a topology, average connected clusterheads value falls sharply up to K=3 and then it becomes constant. This is good for the network; if we have a very high value of average connected clusterheads, leads to congestion in the network and reduces the performance. The constant value of average connected clusterheads at and after K=3 denotes that number of clusters and total connectivity does not change after K=3 as it achieves a constant value.

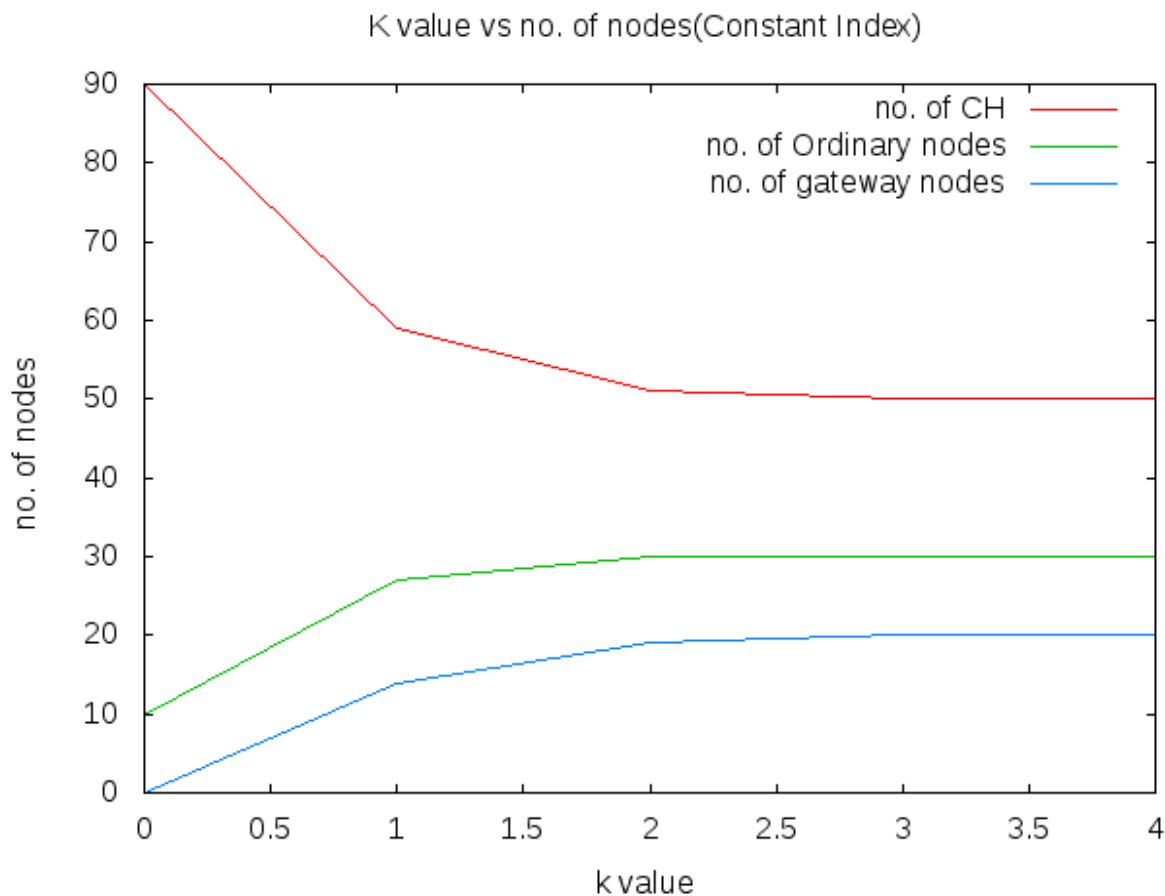

**Fig 5.2.6: Represents the simulation results for 3 different randomly generated topologies with N=100 and varying K value. Topology consists of X, Y and Z co-ordinates of the nodes and N (no. Of nodes) =100.**

Terminologies:
*K value:* This defines that a node which is at most K hops away from the clusterheads can be connected to the Cluster.

Explanation:
With different topologies where initial position varies with variable N value, as we increase the K value in a topology, number of clusterheads decreases and gateway along with ordinary nodes increases. This leads the network to efficiently tackle the congestion at the clusterheads and distributing the load to gateway and ordinary nodes.

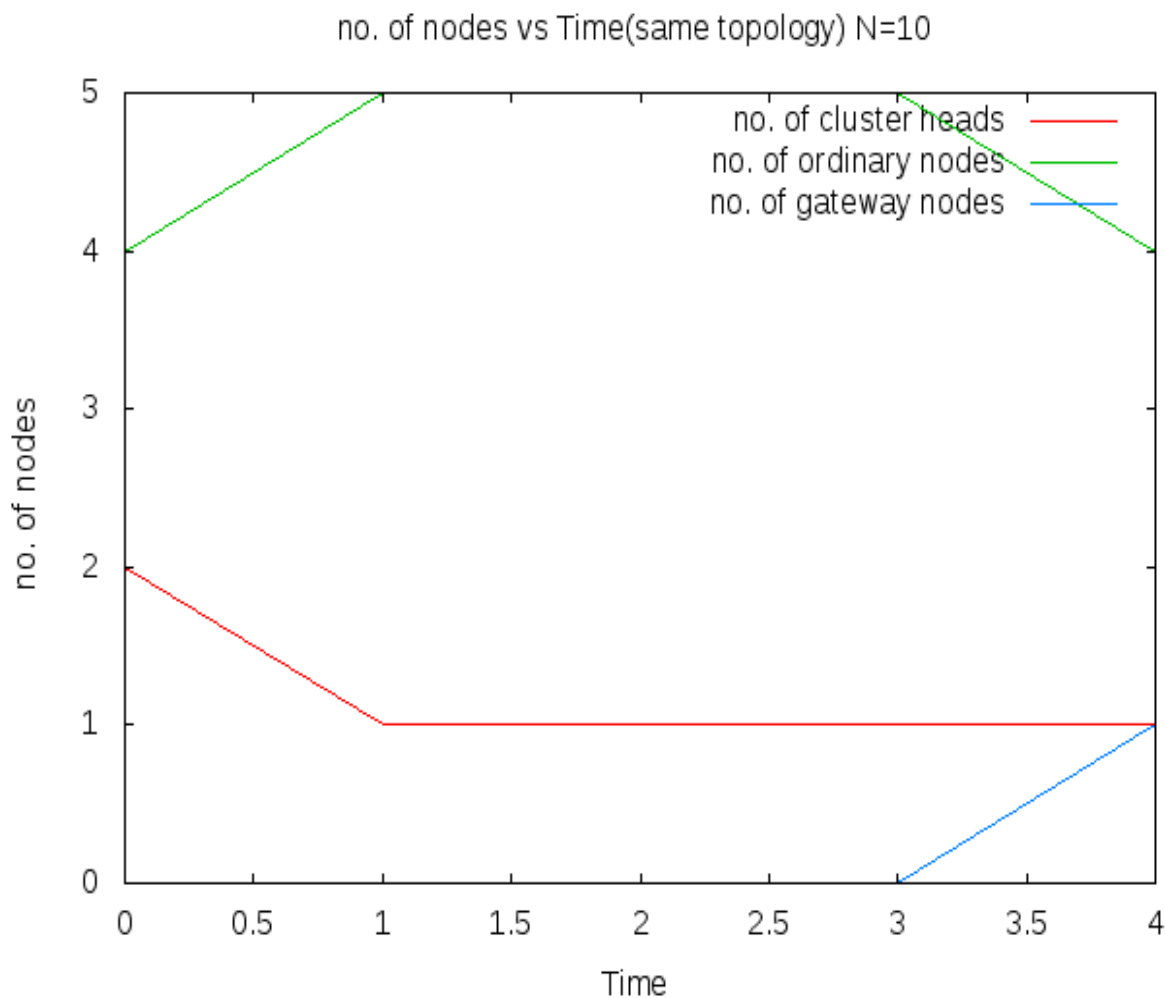

**Fig 5.2.7: Represents the simulation results for 5 instances of a randomly generated topology and having a particular destination, speed and assigned time to reach that destination. Topology consists of X,Y and Z co-ordinates of the nodes and N(no. Of nodes) are 10.**

Explanation:

The simulation results show that for a 10 nodes network having different speed and a final destination at a particular instant of time, our algorithm produces different results for different cases. This specifies in MANET since the nodes are mobile, regular re-run of the clustering algorithm leads to the maintenance feature of our algorithm. This helps to remove the inactive nodes and tackle the mobility. In the above graph we can see the varying no. Of gateway nodes from 0-1 ,ordinary nodes and clusterheads because of the speed of individual nodes which at times doesn't let the nodes to behave as a clusterheads as we have considered a single isolated node as Ordinary node.

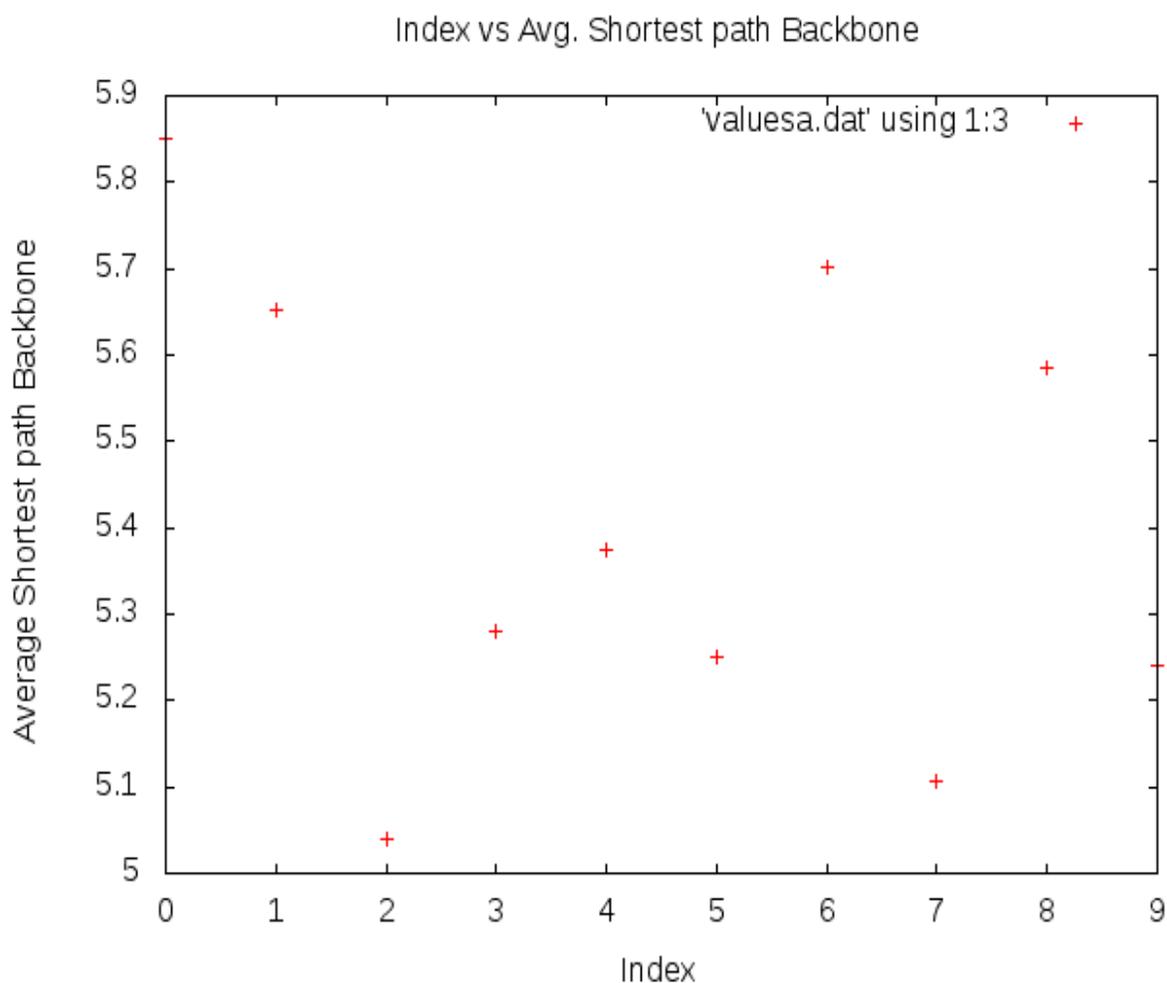

**Fig 5.2.8: Represents the simulation results for 10 different randomly generated topologies of different sizes. Each topology consists of X, Y and Z co-ordinates of the nodes and N(no. Of nodes) is 100**.

Terminology:

*Backbone:* It is a set of dominating nodes .In other words, all the clusterheads in a hierarchical network form a backbone.

*Average shortest path backbone:* This is the average shortest path between the backbone nodes (constituents).

Explanation:

With different topologies where initial position varies with variable N value, as we conduct the simulation for different topologies it can be found that the average shortest path of backbone doesn't have any jumps at all. It has almost constant values which vary from 5.1 to 5.9 in a 100 nodes network.

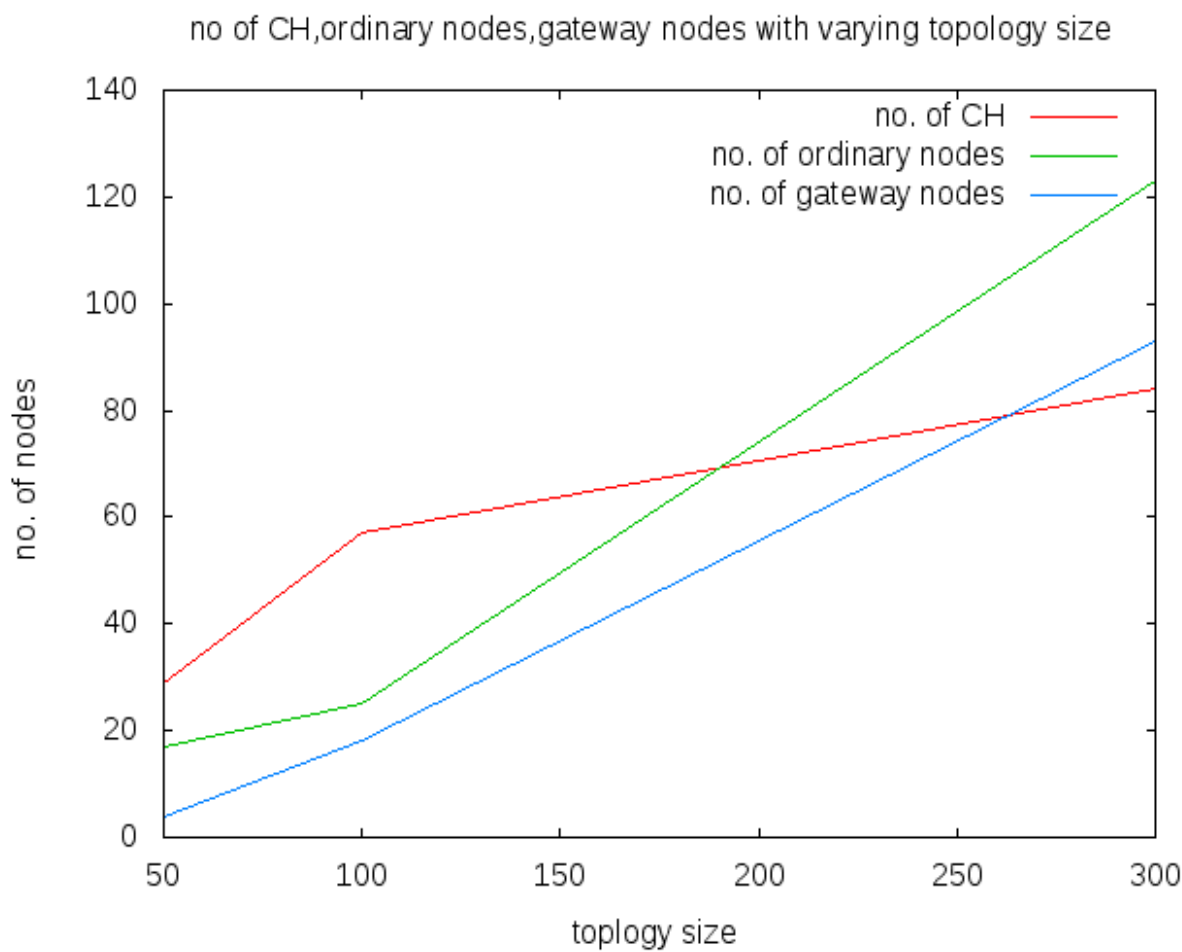

**Fig 5.2.9: Represents the simulation results for 10 different randomly generated topologies of different sizes. Each topology consists of X, Y and Z co-ordinates of the nodes and N (no. Of nodes) is 50,100 and 300.**

Explanation:
With different topologies where initial position varies with variable N value, As we conduct the simulation for different topologies it can be found that with the increase in no. of nodes in a topology number of clusterheads and gateway nodes increases. But in N=3 we have seen a special characteristics of our algorithm which implies that Number of gateway nodes exceeds the number of clusterheads. This is a good sign as too many nodes in a dominating set is also dangerous for the network breakdown and distributing the traffic among the gateway nodes is a good choice.

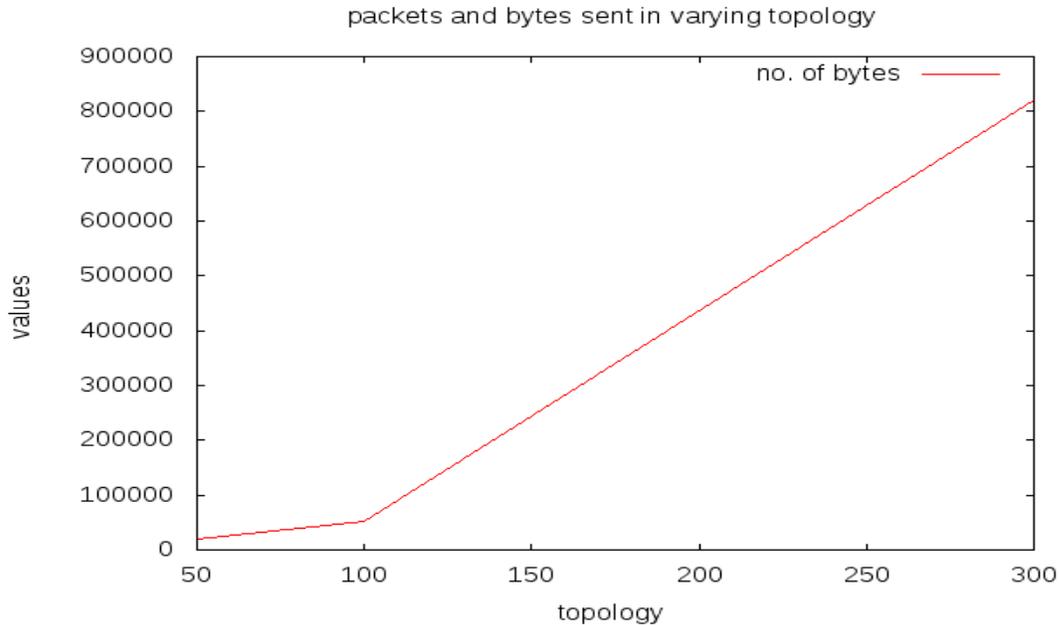

**Fig 5.2.10: Represents the simulation results for 3 different randomly generated topologies of different sizes. Each topology consists of X,Y and Z coordinates of the nodes and N(no. Of nodes) is 50,100 and 300 in the above scenario.**

Terminologies:
Bytes: Amount of information transferred among the nodes.
Explanation:
With different topologies where initial position varies with variable N value, as we increase the number of nodes in a topology, number of bytes transferred also increases. The graph is a linear instead of an exponential one, this denotes the algorithm is scalable on the basis of data. There is not a very steep rise in number of packets as the number of node increases.

# 6.  CONCLUSION AND FUTURE WORK

MANETs have attracted a lot of research addressing all kinds of issues related to them. Dynamic routing plays an important role in the performance of a MANET, and research associated with routing is always a focus. One related topic is the scalability of a routing protocol. Since a large-scale MANET cannot guarantee performance with a flat structure, many cluster hierarchy algorithms have been proposed to solve the scalability issue.

In this thesis, we provided with the fundamental concepts about clustering, including the definition of cluster and clustering, the necessity of clustering for a large dynamic MANET, and the side effects and cost of clustering. Then we presented the related research work which has already been done in this domain. In the same section, we classified proposed clustering schemes into categories based on their main objectives. We discussed each clustering scheme in terms of objective, mechanism, performance, and application scenario, and discussed the benefits and drawbacks of the clustering schemes.

Then we presented a novel multi hop, energy aware algorithm which is scalable and adaptable for various mobility conditions. Cluster formation phase involves little control message overhead. The simulation results and performance evaluation show that the number of clusters formed is proportional to number of nodes in MANET and as the node density increases in the MANET, the number of members in the cluster increases due to k-hop reach-ability criteria . The clusters formed also do not show much variation with mobility.  Apart from this every node is capable of computing its priority based on factors such as battery power, mobility, memory capability and processing power which do not vary rapidly with respect to time. This makes the algorithm energy aware and scalable.

The problem of cluster maintenance is not taken care by the proposed algorithm. The algorithm heavily depends on expensive reclustering operation and the number of clusters varies across the duration of simulation. Therefore an effective cluster maintenance scheme can be incorporated with the proposed clustering algorithm that can be taken up as the future work.